\documentclass[aip,graphicx,rsi,amsmath,amssymp, reprint]{revtex4-1}
\usepackage{graphicx}
\begin{document}
\hfill {\today}

\title{Bent crystal spectrometer for both frequency and wavenumber resolved x-ray scattering at a seeded free-electron laser}

\author{Ulf Zastrau}
\email[]{ulf.zastrau@uni-jena.de}
\affiliation{Institute of Optics and Quantum Electronics, Friedrich-Schiller University Jena, Max-Wien-Platz 1, 07743 Jena, Germany}
\affiliation{Stanford Linear Accelerator Center (SLAC), 2575 Sand Hill Road, Menlo Park, CA 94025 U.S.A.}

\author{Luke B. Fletcher}
\affiliation{Stanford Linear Accelerator Center (SLAC), 2575 Sand Hill Road, Menlo Park, CA 94025 U.S.A.}

\author{Eckhart F\"orster}
\affiliation{Institute of Optics and Quantum Electronics, Friedrich-Schiller University Jena, Max-Wien-Platz 1, 07743 Jena, Germany}
\affiliation{Helmholtz Institute, Fr\"obelstieg 3, 07743 Jena, Germany}

\author{Eric Ch. Galtier}
\affiliation{Stanford Linear Accelerator Center (SLAC), 2575 Sand Hill Road, Menlo Park, CA 94025 U.S.A.}

\author{Eliseo Gamboa}
\affiliation{Stanford Linear Accelerator Center (SLAC), 2575 Sand Hill Road, Menlo Park, CA 94025 U.S.A.}

\author{Siegfried H. Glenzer}
\affiliation{Stanford Linear Accelerator Center (SLAC), 2575 Sand Hill Road, Menlo Park, CA 94025 U.S.A.}

\author{Philipp Heimann}
\affiliation{Stanford Linear Accelerator Center (SLAC), 2575 Sand Hill Road, Menlo Park, CA 94025 U.S.A.}

\author{Heike Marschner}
\affiliation{Institute of Optics and Quantum Electronics, Friedrich-Schiller University Jena, Max-Wien-Platz 1, 07743 Jena, Germany}

\author{Bob Nagler}
\affiliation{Stanford Linear Accelerator Center (SLAC), 2575 Sand Hill Road, Menlo Park, CA 94025 U.S.A.}

\author{Andreas Schropp}
\affiliation{Stanford Linear Accelerator Center (SLAC), 2575 Sand Hill Road, Menlo Park, CA 94025 U.S.A.}

\author{Ortrud Wehrhan}
\affiliation{Institute of Optics and Quantum Electronics, Friedrich-Schiller University Jena, Max-Wien-Platz 1, 07743 Jena, Germany}

\author{Hae Ja Lee}
\affiliation{Stanford Linear Accelerator Center (SLAC), 2575 Sand Hill Road, Menlo Park, CA 94025 U.S.A.}

\begin{abstract}
We present a cylindrically curved GaAs x-ray spectrometer with energy resolution $\Delta E/E = 1.1\cdot 10^{-4}$ and wave-number resolution of $\Delta k/k = 3\cdot 10^{-3}$, allowing plasmon scattering at the resolution limits of the Linac Coherent Light Source (LCLS) x-ray free-electron laser. It spans scattering wavenumbers of 3.6 to $5.2/$\AA\ in 100 separate bins, with only 0.34\% wavenumber blurring. The dispersion of 0.418~eV/$13.5\,\mu$m agrees with predictions within 1.3\%. The reflection homogeneity over the entire wavenumber range was measured and used to normalize the amplitude of scattering spectra. The proposed spectrometer is superior to a mosaic HAPG spectrometer when the energy resolution needs to be comparable to the LCLS seeded bandwidth of 1~eV and a significant range of wavenumbers must be covered in one exposure. 
\end{abstract}


\maketitle

\section{Introduction}

Spectrally resolved x-ray scattering can be used as a novel probing technique to directly measure dense plasma conditions of isochorically heated or laser-compressed solids at mega bar pressures. The ultrafast time resolution provided by the Linac Coherent Light Source (LCLS) x-ray laser allows for studies of high-pressure phase transitions\,\cite{coppari2013,lindenberg2000time}, observations of novel structural properties~\cite{Fortmann12}, or direct measurements of material strain rates. Moreover, knowledge of dense plasma conditions are important for warm dense matter studies and potential applications related to particle acceleration~\cite{hegelich2006,sentoku2011}, inertial confinement fusion~\cite{Lindl04,Glenzer12}, and laboratory astrophysics~\cite{blandford1987particle,kugland2012self}.

Unprecedented experimental capabilities have recently become available to accurately explore extreme matter conditions~\cite{GlenzerRev} with both the LCLS x-ray laser~\cite{emma2010first} and the commissioning of the Matter in Extreme Conditions end station (MEC). This end station is equipped with two nanosecond laser beams, at 2.5~GW, that can drive material into extreme matter conditions by launching shock waves that propagate through solid targets. Under these conditions, spectrally resolved x-ray Thomson scattering measurements in the non-collective (backward) scattering regime can provide information of the microscopic physics by measuring the free electron distribution function. In addition, measurements in the forward scattering regime, collective electron oscillations (plasmons)~\cite{glenzer:065002,Kritcher08,Neumayer10} can simultaneously be observed. The plasmon scattering spectrum is of fundamental interest because it holds promise to determine plasma parameters and the physical properties from first principles~\cite{doppner2009temperature}. This is particularly relevant for plasmas at and above solid density where the material is often strongly coupled and standard theoretical approximations that have been developed for solids, or ideal plasmas, are not applicable. Previous x-ray Thomson scattering studies that use laser-generated x-ray sources in the collective regime have often been insufficient at resolving low-density plasmas due to the large bandwidth of the input source spectrum~\cite{kritcher2009measurements}.

The LCLS beam, in seeded mode operation, delivers approximately $10^{12}$ x-ray photons in a micron-scale focal spot allowing measurements with high spectral resolution of $\Delta E/E = 10^{-4}$, high wave-number resolution of $\Delta k/k = 10^{-2}$, and high temporal resolution of $20-50\,$fs.  Consequently, by employing highly efficient curved crystal spectrometers~\cite{Zastrau2012,Zastrau2013}, the plasmon spectrum can be observed and resolved in a single x-ray pulse. 

\section{Principle}

\begin{figure}
 		\includegraphics[width=0.48\textwidth]{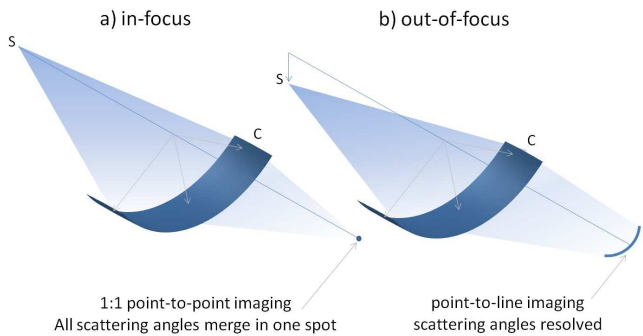}
 		\caption{Schematics of the spectrometer principle: In the classical von-H{\'a}mos geometry (a), the source of scattered x-rays $S$ lies on the cylinder axis of the crystal $C$, giving rise to a 1:1 imaging onto the axis. But when the source $S$ is positioned below the axis (b) the setup is \textit{out of focus}, resulting in a curved line in the detector plane on-axis. In the latter case different scattering angles can be resolved.}
 		\label{fig:principle}
 \end{figure}

Until recently, x-ray scattering experiments at the LCLS were only possible in self-amplified spontaneous emission (SASE) mode, having a spectral bandwidth of $\sim 3$\%, e.g., $\sim 20~$eV at 8~keV photon energy. Mosaic cylindrical crystals from highly oriented pyrolytic graphite (HOPG)~\cite{Zastrau2012} in von-H{\'a}mos geometry~\cite{Hamos32} have been successfully employed in novel scattering experiments~\cite{brown2014evidence}, since their energy resolution matches the SASE bandwidth. When operated in seeded mode, crystals from highly-annealed pyrolytic graphite (HAPG)~\cite{Zastrau2013} have shown to enable plasmon scattering~\cite{fletcher2013plasmon} at the LCLS, but their energy resolution of $\Delta E = 9\,$eV is still an order of magnitude worse than the seeded LCLS bandwidth. A crystal with matching energy resolution of $\leq 1~$eV at 8~keV would allow the determination of the plasmon width, and hence the collisionality via plasmon damping~\cite{Neumayer10,doppner2010x}. Instead of using mosaic crystals, the use of perfect crystals allows achieving these energy resolutions.

The efficiency of a cylindrically bent crystal in von-H{\'a}mos geometry is determined by the reflection curve in the dispersive plane, and by its collection width. Wide crystals are efficient but cover a significant solid angle. In the standard geometry, point-to-point focusing is obtained with 1:1 magnification (cf. fig.~\ref{fig:principle}), and x-rays with different scattering wavenumbers can no more be distinguished in the focus. This becomes critical when the wavenumber range comprises Bragg reflections. These have intensities orders or magnitude larger than plasmon scattering and completely outshining the latter, potentially saturating the detector in a single shot. 

Here, we take advantage of the capability of perfect crystals to act as high-quality optics, since the reflected beams are only broadened by the rocking curve width of a few seconds of arc. When the geometry is set slightly \textit{out-of-focus} as depicted in fig.~\ref{fig:principle}, the x-rays will form a hyperbolic line-focus with the different wavenumbers being spatially resolved.

\section{Design of the spectrometer}

\begin{figure}
 		\includegraphics[width=0.48\textwidth]{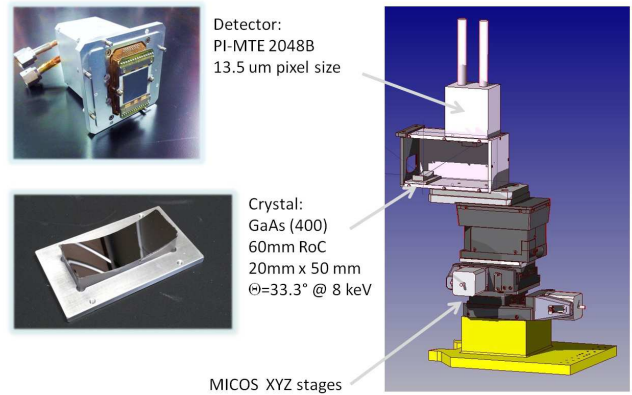}
 		\caption{Components of the GaAs spectrometer. The CCD detector is mounted onto a light-tight housing, which contains the GaAs 400 crystal right after a removable window slider. This window is equipped with Al-coated mylar foil to suppress visible light. The housing is mounted on a motorized stage to allow three-dimensional positioning with respect to the source point.}
 		\label{fig:components}
 \end{figure}

As key component of such a spectrometer, we have chosen a crystal of GaAs in orientation (100). It has the same lattice parameters as Ge, but more practical experience in grinding it down to $\sim 60~\mu$m thickness is available due to its use in the semiconductor industry. The thin crystal was bent and glued to a toroidal glass lens ($20 \times 50\,$mm$^2$) with a radius of curvature (RoC) of $60\,$mm using strain-free glue (cf. fig.~\ref{fig:components}). A CCD detector (Princeton Instruments PI-MTE 2048B, pixel pitch $13.5\,\mu$m) is mounted on top of a light-tight aluminum housing, such that the chip is oriented parallel to the cylinder axis of the crystal. The x-rays enter the housing though a window consisting of three layers of $1.5\,\mu$m mylar foil flash-coated with 100~nm of aluminum on both sides (transmission $T=0.995$ at 8~keV). The entire setup is designed to operate under high vacuum conditions (pressure $<10^{-4}\,$mbar) when the CCD is cooled.

For reflection 400 at 8~keV, the Bragg angle is $\Theta_B=33.25^\circ$, yielding a focal length $f$ (distance from source to crystal center) of $109\,$mm. The entire box is mounted on a vacuum-compatible motorized x-y-z translation stage to allow in-vacuum positioning of the spectrometer with respect to the source point.

 \begin{figure}
 		\includegraphics[trim=0cm 0cm 0cm 0cm, width=0.45\textwidth]{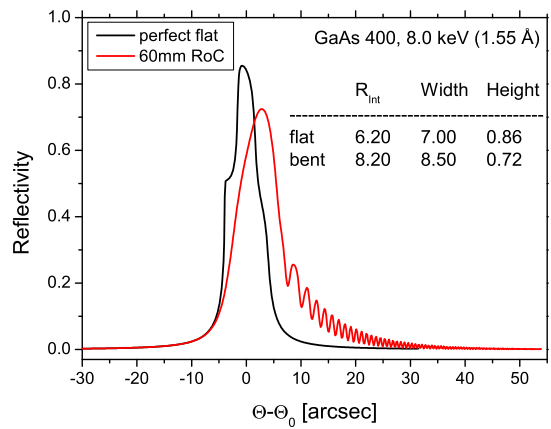}
 		\caption{Rocking curves of flat and bent GaAs 400 crystals, calculated with the code DIXI~\cite{Hoelzer98} for the experimental conditions under consideration. The values in the inset are given in seconds of arc. The integrated reflectivity $R_{\rm int}$ is expected to increase by 32\% from $6.2$  to $8.2\,$arcsec when the perfect flat crystal is bent to $60\,$mm radius of curvature (RoC).}
 		\label{fig:rockingcurves}
 \end{figure}

The resolving power of a perfect crystal is determined by its rocking curve width. Rocking curves of flat and bent GaAs 400 crystals have been calculated with the code DIXI~\cite{Hoelzer98} for the experimental conditions under consideration. As can be seen from fig.~\ref{fig:rockingcurves}, when a perfect flat crystal is bent to $60\,$mm radius of curvature (RoC), the rocking curve obtains an asymmetric shape with wiggles, increasing its width. This effect is due to the gradient in lattice spacing for different crystal depths and has been experimentally confirmed~\cite{Uschmann93}. The integrated reflectivity $R_{\rm int}$ is expected to increase by 32\% from $6.2$  to $8.2\,$arcsec. The divergence due to the rocking curve width of 8.5~arcsec ($\Delta \Theta =41\,\mu$rad) results in a width of $9\,\mu$m when the rays from an ideal point source have propagated twice the focal distance $2f = 218\,$mm. This leaves a footprint of $16\,\mu$m on the detector, which matches the typical pixel pitch of the employed x-ray CCD ($13.5\,\mu$m). The expected energy resolution is $\Delta E/E = \Delta \Theta / \tan \Theta_B = 6.2\times 10^{-5}$, or  $\Delta E = 0.5\,$eV at $E=8\,$keV.


\section{Performance at the LCLS}

\begin{figure}
 		\includegraphics[width=0.48\textwidth]{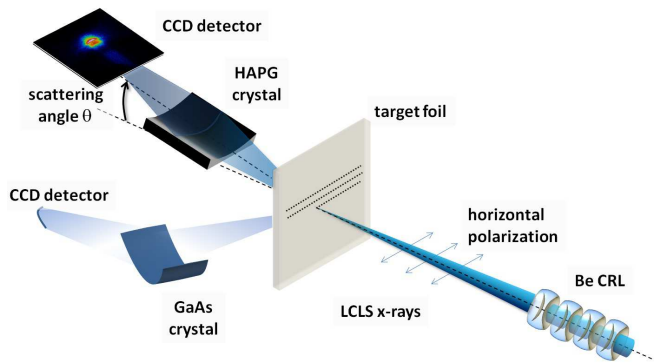}
 		\caption{Experimental setup at the Matter in Extreme Conditions (MEC) instrument at the LCLS. Horizontally polarized 8~keV photons are focused onto a target foil using Be compound refractive lenses (CRL), giving rise to x-ray scattering. At the foil rear side, a HAPG and a GaAs spectrometer are positioned at effective scattering angles of $\sim 25^\circ$ and $\sim 60^\circ$, respectively.}
 		\label{fig:LCLS_setup}
 \end{figure}

For comparison of the performance of the new spectrometer with a standard mosaic crystal spectrometer under experimental conditions at the LCLS, we have mounted it together with a HAPG spectrometer~\cite{Zastrau2013} in the geometry depicted in fig.~\ref{fig:LCLS_setup} in the vacuum chamber at the MEC end station. 

The HAPG spectrometer employs a $30\times 32\,$mm$^2$ large crystal with a radius of $51.7\,$mm (details are given by Zastrau et al.~\cite{Zastrau2013}). The lattice spacing is $6.708\,$\AA, which results in a Bragg angle of $\Theta_B=13.3^\circ$ at 8~keV. It is mounted exactly in forward scattering direction above the LCLS beam. Its scattering angle is thus simply given by the angle $\Theta$ in the vertical plane. On the other hand, the position of the GaAs spectrometer makes it necessary to calculate the effective scattering angles and the resulting wavenumber range.

\subsection{Effective scattering angle}

In the following an analytic expression for the effective scattering angle of an arbitrarily positioned cylindrical crystal spectrometer will be derived. 

\begin{figure*}
\centering
\includegraphics[width=0.9\textwidth]{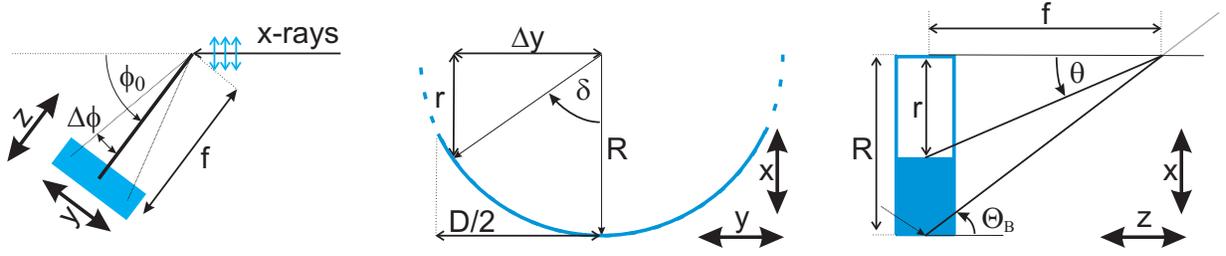}
\caption{Coordinates and conventions for the analytic expression. Left: Let a cylindrical crystal (blue) be positioned  at angle $\phi_0$ in the polarization plane of the x-rays, covering an angular range of $\pm \Delta \phi$. Cartesian coordinates are chosen with respect to the spectrometer, where the $x,y$-plane is perpendicular to the cylinder axis (middle), and the $x,z$-plane is the dispersion plane (right).} 
\label{fig:coordinates} 
\end{figure*}

As shown in the schematic in fig.~\ref{fig:coordinates}, Cartesian coordinates are chosen with respect to the von-H{\'a}mos spectrometer, where the $x,y$-plane is perpendicular to the cylinder axis (center panel), and the $x,z$-plane is the dispersion plane (right panel). The cylinder axis of the crystal (blue) is positioned at angle $\phi_0$ in the polarization plane of the x-rays, covering an angular range of $\pm \Delta \phi$. For the sake of simplicity, let the crystal be positioned at its Bragg angle $\Theta_B$ perpendicular to this plane, such that the crystal cylinder axis with radius $R$ lies in the same plane as the $y,z$-plane. We will generalize this constrain later.   

The effective scattering angle $\theta_\mathrm{eff}$ is given by the dot product of $\phi=\phi_0+\Delta\phi$ and $\theta$. The $\phi$-contribution is simply the angle between the spectrometer axis and the incoming x-rays in the $y,z$-plane. In general, $\theta \leq \Theta_B$ due to the curvature of the crystal.

From the identity $f \cdot \tan \Delta\phi = \Delta y = R\cdot \cos \delta$, we derive that the height displacement due to the curvature of the crystal amounts to $r = R\cdot \sin \delta$, or more generally
\begin{eqnarray}\label{r_of_phi}
\begin{split}
r(\Delta \phi) = R\cdot\underbrace{\sin\left[\arccos \left( \frac{f}{R} \tan \Delta \phi \right) \right]}_{\sin(\arccos x) = \sqrt{1-x^2} } = \\ 
= R\cdot\sqrt{1-\left( \frac{f}{R} \tan \Delta \phi\right)^2}
\end{split}
\end{eqnarray}
The expression $r/f=\tan \theta$ assumes $R/f=\tan \Theta_B$ for $\Delta\phi=0$, which is also the solution for a flat crystal. By substituting $R/f$ by $\tan \Theta_B$ in Eq.(\ref{r_of_phi}), and writing $\theta$ as function of $r(\Delta\phi)$, we obtain
\begin{eqnarray}\label{theta_of_phi}
\theta(\Delta \phi) = \arctan \left[ \tan \Theta_B \cdot \sqrt{1-\left( \frac{\tan \Delta \phi}{\tan \Theta_B}\right)^2}~ \right].
\end{eqnarray}
We note that $\theta(0) = \Theta_B$ for the center of the crystal and $\theta(\Theta_B) = 0$. The effective scattering angle is thus given by
\begin{eqnarray}\label{theta_eff}
\Theta_\mathrm{eff}(\Theta_B, \phi, \Delta\phi)= \arccos \left[ \cos \theta(\Delta\phi) \cdot \cos (\phi+\Delta\phi)\right].
\end{eqnarray}

For the GaAs spectrometer as employed in this work, $\Theta_B=33.25^\circ$, $\phi=53^\circ$ and $\Delta\phi=\pm 15^\circ$, hence we obtain angular limits of $47.5^\circ$ and $71.3^\circ$ with a central scattering angle of $59.8^\circ$.

For the HAPG spectrometer we have to consider the general case of the cylinder axis being at arbitrary angle $\theta_A$ out of the $x,z$-plane. Eqn.(\ref{r_of_phi}) and eqn.(\ref{theta_of_phi}) are not affected since they only describe a relative change of $r$ and $\theta$ in the coordinate system of the spectrometer. Hence, $\theta_A$ can simply be accounted for in Eq.(\ref{theta_eff}:
\begin{eqnarray}
\begin{split}\label{theta_eff2}
\Theta_\mathrm{eff}(\Theta_B, \theta_A; \phi, \Delta\phi)= \\
= \arccos \left[ \cos \left( \theta_A + \theta(\Delta\phi) \right) \cdot \cos (\phi+\Delta\phi)\right].
\end{split}\end{eqnarray}

For HAPG, we find at $\Theta_B=13.3^\circ$, horizontal cylinder axis angle $\theta_A=11.65^\circ$, vertical scattering angle $\phi=0^\circ$, and $\Delta\phi=\pm 4^\circ$ symmetric angular limits of $24.74^\circ$ with a central scattering angle of $\Theta_B+\theta_A=25^\circ$ ($\Delta\Theta \sim 0.26^\circ$). We note that angular spread over which the scattered signal is obtained is given by the mosaicity of the crystal of $\omega = 0.14^\circ$ and the detected spectral range, which enters through the dispersion $\Delta\Theta_{\rm eff}/\Delta\lambda$. For a spectral range of $\pm 35$~eV, typical for plasmons, we find $\Delta\Theta_{\rm eff} \sim 0.11^\circ$, corresponding to a $2\,$mm wide strip on the crystal.

\subsection{Wavevector range and blurring}

Different from GaAs, the mosaic HAPG merges all collected x-rays from different scattering angles into one unresolved focus. The total angular blurring is determined by the width of the crystal, the mosaicity, and the spectral range (see previous discussion), and results in a total $\Delta\Theta_{\rm eff} = 0.5^\circ$ ($9\,$mrad). The scattering wavenumber $k=\mid \vec{k} \mid$ is defined as~\cite{GlenzerRev}
\begin{eqnarray}\label{wavevector_HAPG}
k= \frac{4\pi}{hc} E_0 \sin \frac{\Theta}{2} = 1.013  \frac{E_0[\mathrm{keV}]}{\rm \AA} \sin \frac{\theta}{2} =  1.9/{\rm \AA}
\end{eqnarray}
Here we use $E_0=8\,$keV and $\Theta=25^\circ$. The wavenumber blurring $\Delta k$ is related to the angular blurring $\Delta \theta$ by 
\begin{eqnarray}
\Delta k = 0.506 \frac{E_0[\mathrm{keV}]}{\rm \AA} \cos \frac{\Theta}{2} \cdot \Delta\Theta  =  0.035/{\rm \AA}
\end{eqnarray}
This is a wavenumber blurring of $\sim 1.8$\%.

The focus of the GaAs spectrometer was chosen to be about 2~mm below the cylinder axis, resulting in a 100~pixel wide profile ($1.35\,$mm length) on the detector. On average, each pixel covers an angular range of  only $\Delta \Theta_{\rm eff} \sim 0.24^\circ$ ($4.2\,$mrad). Here, 
\begin{eqnarray}\label{wavevector_GaAs}
k = 3.6~ ... ~5.2/{\rm \AA}, {\rm ~but~} \Delta k \sim 0.015/{\rm \AA},
\end{eqnarray} 
which is only $0.34$\%. We note that, eventhough the wavenumber range of the GaAs spectrometer is 46~times larger, the wavenumber blurring per bin is 5~times improved.

\subsection{Energy resolution and dispersion}

In order to calibrate the photon energy dispersion and spectral resolution, we analyze the K$\alpha$ fluorescence from a 25$\,\mu$m thin Cu foil, located at $E(K\alpha_1)=8048~$eV and $E(K\alpha_2)=8028~$eV. To generate fluorescence, the LCLS was tuned to a photon energy of 9000~eV, i.e. above the K-absorption edge of Cu at $E_{\rm K-edge} = 8980~$eV, focused to a spot of $1\,\mu$m diameter to minimize source broadening, and attenuated to 1\% to prevent isochoric x-ray heating. 

  \begin{figure}
 		\includegraphics[trim=0cm 0cm 0cm 0cm, width=0.45\textwidth]{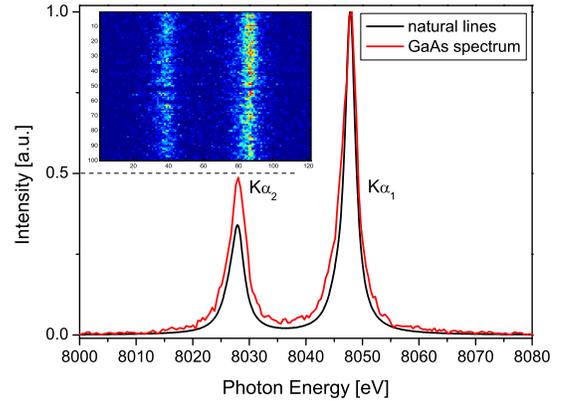}
 		\caption{Single-shot Cu K$\alpha$ fluorescence as measured with the GaAs spectrometer, together with the published natural line shapes~\cite{Krause1979,Holzer97} for comparison. The inset shows the raw data after taking out the curvature.}
 		\label{fig:CuKa}
 \end{figure}
 
Figure~\ref{fig:CuKa} shows a single-shot Cu K$\alpha$ fluorescence recorded by the GaAs spectrometer.  For the raw data shown in the inset, the curvature was taken out by shifting the pixels in the spectral direction according to a hyperbolic function, corresponding to the intersection of the x-ray cone with the detector plane. Also shown are  the published natural line shapes~\cite{Krause1979,Holzer97} for comparison. We found that the best agreement is achieved when applying a linear dispersion of $0.418\,$eV/pixel, which is only $1.3$\% different from the theoretical dispersion of $0.4126\,$eV/pixel (based on geometrical calculations).

The full width at half maximum (FWHM) of the Cu K$\alpha_1$ line from the literature~~\cite{Krause1979,Holzer97} amounts to 2.24~eV whereas our measured value exhibits a FWHM of 1.13~eV. Since the widths of two convoluted Lorentzians are additive, we derive an energy resolution of about 0.89~eV at 8048~eV, or a resolving power of $\Delta E / E \sim 10^{-4}$. This value is about 2~times larger than theoretically expected from the rocking curve width, but fully explainable by the sampling of the x-ray focus by the CCD pixels and blooming. It perfectly matches the bandwidth of the seeded LCLS.

\subsection{Reflection homogeneity}

\begin{figure}
 		\includegraphics[trim=0cm 0cm 0cm 0cm, width=0.49\textwidth]{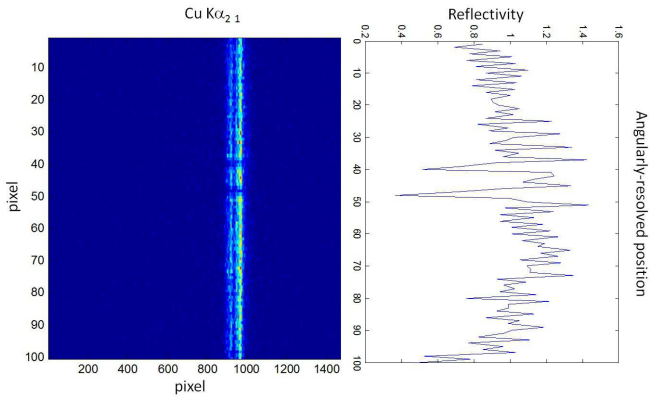}
 		\caption{The left image shows the Cu K$\alpha$ fluorescence as measured with the GaAs spectrometer, which is angularly resolved from top to bottom. The right graph shows the projection along the spectral domain (left-right), allowing to infer the changes in reflectivity, which range between 0.3 and 1.4. This data can be used to normalize the amplitude of scattering spectra. }
 		\label{fig:homogeinity}
 \end{figure}

Different from x-ray scattering, the K$\alpha$ fluorescence has no angular dependence other than reabsorption in the Cu foil itself. Therefore, the angularly-resolved intensity distribution on the detector is a measure for the reflection homogeneity of the crystal. Figure~\ref{fig:homogeinity} shows the Cu K$\alpha$ doublet emission, accumulated over five exposures. When projecting it in the spectral direction (right part of fig.~\ref{fig:homogeinity}) the resulting profile shows the reflectivity for each scattering angle, which ranges between $0.3$ and $1.4$. These variations can be due to defect structures in the crystal material itself, areas of imperfect curvature, or irregularities in the CCD detector sensitivity. This data can be used to normalize the amplitude of scattering spectra.

\subsection{Relative throughput}

Since both the GaAs and the HAPG spectrometer employ x-ray CCD detectors (PI-MTE with BN chip), we may compare their relative photon counts for the Cu K$\alpha$ fluorescence after treating the reabsorption in the Cu foil properly. Here, we assume the dominant source of fluorescence to be located at the absorption length for 9~keV x-rays, $l_{\rm abs} = 4\,\mu$m below the front surface of the $25\,\mu$m thin Cu foil. From the effective observation angles for HAPG ($25^\circ$) and GaAs ($60^\circ$), the generated Cu K$\alpha$ x-rays travel $23\,\mu$m and $43\,\mu$m through Cu, respectively, before reaching the foil rear surface. This implies transmissions of $T=35$\% and $T=14$\%, respectively. Including slight differences in filter transmission in front of either spectrometer, we find that, in total, the HAPG spectrometer sees $1.9~$times more fluorescence photons as compared to the GaAs spectrometer, which has to be corrected for.

From the experiment, after correcting for filters and reabsorption, we find that the HAPG CCD detects 12~times more Cu K$\alpha$ photons. But the HAPG spectrometer has both less energy and wavenumber resolution, which results in less detected photons per spectral and momentum-transfer element, as shown in tab.~\ref{tab:eff}. Already the 10-fold higher spectral resolution of GaAs almost compensates for the higher efficiency of HAPG. When high wavenumber resolution is needed, i.e. to discriminate plasmon scattering from nearby Bragg peaks, the differential signal on the GaAs spectrometer is twofold higher.

\begin{table} \centering
\begin{tabular}{l||c|c|c|c|}
crystal & relative  & spectral & momentum & differential \\
~       & signal & element  & element  & signal \\
~       &  $S$     & $\Delta E$  [eV]& $\Delta k$ [${\rm \AA}^{-1}$] & $S/\Delta E / \Delta k$ \\
	\hline 
HAPG & 12$\times$  & 9    & 0.035 & 38\\
GaAs & 1$\times$   & 0.89 & 0.015 & 75\\
    \hline
ratio & 12:1 & 10:1 & 2.3:1 & 1:2 \\
    \hline
\end{tabular}
\caption{Comparison of relative efficiency, energy, and momentum elements for both spectrometers under consideration.}\label{tab:eff}
\end{table}

\subsection{Proof-of-principle experiment}

\begin{figure}
 		\includegraphics[trim=0cm 0cm 0cm 0cm, width=0.49\textwidth]{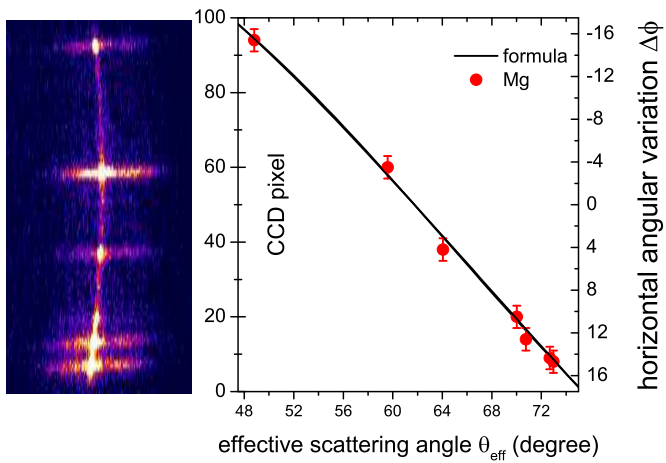}
 		\caption{Left: Raw image of 150 accumulated scatter events from a Mg foil at 8~keV. Right: The CCD pixel position of the Bragg peaks (y-axis) correlated with their tabulated values (x-axis). The data is in good agreement with the effective scattering calculated via eqn.~\ref{theta_eff} (solid line).}
 		\label{fig:Mg_scatter}
 \end{figure}

As a proof of principle, the LCLS was operated at 8~keV photon energy in seeded beam mode, yielding a bandwidth of 1~eV only. As prototypical mid-Z specimen, we chose a polycrystalline foil of Mg, having a hexagonal close-packed structure and lattice constants of $3.2$ and $5.2$\AA, respectively. The x-ray beam was focused to a  spot of $10\,\mu$m onto a $50\,\mu$m thin Mg foil. At a repetition rate of 120~Hz, the foil was constantly moving and 150 consecutive scattering events were recorded. The left part of fig.~\ref{fig:Mg_scatter} shows a raw image, where the x-axis is the spectral direction, and the y-axis are the 100 angularly-resolved pixels. On the right, the CCD pixel position of each Bragg peak is correlated with the tabulated values for a Mg crystal. The solid line shows that this data is in good agreement with the effective scattering as calculated via eqn.~\ref{theta_eff}.

\section{Conclusion}

The LCLS beam, in seeded mode operation, delivers approximately $10^{12}$ x-ray photons in a micron-scale focal spot allowing measurements with high spectral resolution of $\Delta E/E = 10^{-4}$, high wave-number resolution of $\Delta k/k = 10^{-2}$, and high temporal resolution of $20-50\,$fs.  Here we present a spectrometer with energy resolution $\Delta E/E = 1.1\cdot 10^{-4}$ and wave-number resolution of $\Delta k/k = 3\cdot 10^{-3}$, allowing plasmon scattering experiments at the resolution limits of the LCLS.

The spectrometer employs a GaAs crystal in reflection 400 that is bent to 60~mm radius of curvature. It was tested under experimental conditions at the MEC instrument at the LCLS. As detector, an x-ray CCD with $13.5\,\mu$m pixel pitch was employed. In an out-of-focus von-H{\'a}mos geometry, the spectrometer spans scattering wavenumbers of 3.6 to $5.2/$\AA\ in 100 separate bins, with only 0.34\% wavenumber blurring per bin, allowing the discrimination of Thomson scattering in the proximity of Bragg reflections.

Using Cu K$\alpha$ fluorescence, we determined the energy resolution to be 0.89~eV at the Cu K$\alpha_1$ line at 8048~eV. The dispersion of 0.418~eV/pixel is within 1.3\% agreement with the design values. Further, we used the fluorescence light to determine the reflection homogeneity over the entire wavenumber range and found it varying between 0.3 and 1.4, normalized to a mean value of 1. These data can be used to normalize the amplitude of scattering spectra.

In comparison to a simultaneously employed mosaic HAPG spectrometer, we find that the total efficiency is 12~times lower, but is quantitatively compensated by the higher spectral resolution. Further, the wavevector blurring of the HAPG spectrometer is 2.3~times worse, and it covers only one scattering angle at a time. 

We conclude that the proposed spectrometer can be superior to a mosaic HAPG spectrometer when the energy resolution needs to be comparable to the LCLS seeded bandwidth of 1~eV and a significant range of wavenumbers has to be covered in a single shot.

\begin{acknowledgments}
The experiments were performed at the Matter at Extreme Conditions (MEC) instrument of LCLS, supported by the DOE Office of Science, Fusion Energy Science under contract No. SF00515. For the preparation of the crystal, the assistance of the Bundesministerium f{\"u}r Bildung und Forschung within the priority research area FSP 301 FLASH and the Volkswagen Foundation via a Peter-Paul-Ewald Fellowship is acknowledged. This work was partially supported by the DOE Office of Science, Fusion Energy Sciences under FWP 100182. 
\end{acknowledgments}



%

\end{document}